\documentclass[letterpaper, 10pt, conference]{ieeeconf}

\IEEEoverridecommandlockouts

\usepackage{balance}

\usepackage{amsmath,mathtools}  
\usepackage{amsfonts,amssymb}
\usepackage{xspace}    

\usepackage{graphicx}
\graphicspath{{figs/}}

\usepackage{subcaption}

\usepackage{hyperref}
\hypersetup{colorlinks=true, linkcolor= blue, allcolors=blue, citecolor = blue, filecolor=black, urlcolor=blue}

\usepackage{pgfplots}\pgfplotsset{compat=1.9} 
\usepackage{grffile}
\pgfplotsset{compat=newest}  
\usetikzlibrary{plotmarks}
\usetikzlibrary{arrows.meta}
\usepgfplotslibrary{patchplots}

\pgfplotsset{compat=1.15}
\usepackage{mathrsfs}
\usetikzlibrary{arrows}

\usepackage{tabularx}

\usepackage[ruled]{algorithm}
\usepackage[noend]{algpseudocode}

\makeatletter
\newcommand{\multiline}[1]{%
  \begin{tabularx}{\dimexpr\linewidth-\ALG@thistlm}[t]{@{}X@{}}
    #1
  \end{tabularx}
}
\makeatother

\usepackage[per-mode=symbol]{siunitx}

\usepackage{cite}  
\usepackage{array}  
\usepackage{booktabs}           

\newtheorem{theorem}{Theorem}

\newcounter{remark}
\newenvironment{remark}{%
\par\vspace{3pt}\noindent\refstepcounter{remark}\textbf{Remark~\theremark:}}%
{\par\endtrivlist\unskip}

\newcounter{problem}
{\par\endtrivlist\unskip}

\usepackage[extramath,probability,reviewmark]{truonglatexdefs}

\usepackage{stfloats}

\title{\LARGE \bf
A Cooperative Optimal Control Framework for Connected and Automated Vehicles in Mixed Traffic Using Social Value Orientation}
\author{Viet-Anh Le, \emph{IEEE Student Member}, Andreas A. Malikopoulos, \emph{IEEE Senior Member}%
\thanks{
The authors are with the Department of Mechanical Engineering, University of Delaware, Newark, DE 19716 USA. E-mail: \tt\small{vietale@udel.edu}, \tt\small{andreas@udel.edu}.} 
}

\begin{document}

\maketitle
\thispagestyle{empty}
\pagestyle{empty}

\begin{abstract}

In this paper, we develop a socially cooperative optimal control framework to address the motion planning problem for connected and automated vehicles (CAVs) in mixed traffic using social value orientation (SVO) and a potential game approach.
In the proposed framework, we formulate the interaction between a CAV and a human-driven vehicle (HDV) as a simultaneous game where each vehicle minimizes a weighted sum of its egoistic objective and a cooperative objective. 
The SVO angles are used to quantify preferences of the vehicles toward the egoistic and cooperative objectives. 
Using the potential game approach, we propose a single objective function for the optimal control problem whose weighting factors are chosen based on the SVOs of the vehicles. 
We prove that a Nash equilibrium can be obtained by minimizing the proposed objective function.
To estimate the SVO angle of the HDV, we develop a moving horizon estimation algorithm based on maximum entropy inverse reinforcement learning.
The effectiveness of the proposed approach is demonstrated by numerical simulations of a vehicle merging scenario. 
\end{abstract}

\section{Introduction}
\label{sec:intro}

Coordination of connected and automated vehicles (CAVs) has received significant attention over the last years (see \cite{zhao2019enhanced,guanetti2018control,ersal2020connected} for surveys).
In recent work \cite{mahbub2020decentralized,remer2019multi,mahbub2020Automatica-2}, we addressed optimal coordination of CAVs at different traffic scenarios and last-mile delivery applications, and showed potential benefits in reducing traffic congestion, travel time, and energy consumption.
These approaches 
have considered 100\% CAV penetration rate, 
which, however, might not be expected by $2060$ \cite{alessandrini2015automated}.
Therefore, it is highly necessary to investigate motion planning and control strategies for CAVs that perform efficiently in mixed traffic environments. 

A number of research articles have recently focused on developing different control methods for CAVs in mixed traffic scenarios,
\eg 
optimal control \cite{jin2018connected}, 
model predictive control \cite{feng2021robust},
game-theoretic control \cite{chandra2022gameplan},
or learning-based control \cite{bae2020cooperation}. 
Other research efforts \cite{schwarting2019social,buckman2019sharing,ozkan2021socially,larsson2021pro,toghi2021cooperative,crosato2021human} have utilized \emph{social value orientation} (SVO), a concept from psychology that quantifies the level of an agent’s selfishness or altruism, in control development to predict how human drivers will interact and cooperate with others.
The first attempt taking social factors of human drivers into consideration via SVO was made in \cite{schwarting2019social}.
The authors modeled interactions between vehicles as a best-response game in which each agent maximizes its individual utility given optimal actions of other agents, and utilized SVO to better predict human driving intention.
A centralized coordination algorithm was developed by Buckman \etal \cite{buckman2019sharing} to perform socially-compliant navigation at an intersection given the social preferences of the vehicles.
Ozkan and Ma \cite{ozkan2021socially} investigated the impacts of a socially compatible control design in a car-following scenario through four different SVO levels for an automated vehicle.
Larsson \etal \cite{larsson2021pro} developed a pro-social model predictive control algorithm in which SVO was used to derive weighting strategies.
Toghi \etal \cite{toghi2021cooperative} proposed a cooperative sympathetic reward structure using SVO for multi-agent reinforcement learning in autonomous driving.
Crosato \etal \cite{crosato2021human} considered a scenario with pedestrian crossing and utilized SVO to design a reinforcement learning reward function for an automated vehicle to avoid collision with the pedestrian.


Inspired by using SVO to understand the driving preferences of human drivers, in this paper, we develop a \emph{socially cooperative optimal control framework} for a CAV while interacting with a HDV.
Contrary to other efforts that have used the Stackelberg game with a leader-follower structure \cite{schwarting2019social,ozkan2021socially,toghi2021cooperative}, which may require much computation for finding an equilibrium,
our approach models the interaction between the CAV and the HDV as a \emph{simultaneous game}.
In the imposed game, the agents simultaneously take control actions to minimize their objective functions that are given by a weighted sum of their individual (egoistic) objective and a shared (cooperative) one.
The SVO angles are used to quantify how the agent weights its individual objective against the shared objective.
Based on the idea of \emph{potential games} \cite{monderer1996potential}, we then propose an objective function for the control framework which is used to obtain a Nash equilibrium of the imposed game.
The framework is implemented in a \emph{receding horizon control} (or model predictive control) manner for robustness against stochastic human driving behavior.
To estimate the SVO angle of the HDV, we employ \emph{moving horizon estimation} which uses real-time data from both vehicles combined with the \emph{maximum entropy inverse reinforcement learning} (IRL) technique. 
Based on the estimated SVO of the HDV, the SVO of the CAV is chosen to compensate for the level of altruism in CAV\textendash HDV coordination.
Finally, the proposed framework is illustrated and validated by a numerical example of a vehicle merging scenario.

The remainder of this paper is organized as follows.
In Section~\ref{sec:mpc}, we present the proposed socially cooperative optimal control framework for a CAV while interacting with a HDV, and introduce an illustrative example 
at a merging.
In Section~\ref{sec:rhe}, we provide the method for estimating the SVO angle of the HDV, and in Section~\ref{sec:sim}, we demonstrate the performance of the proposed framework by numerical simulations.
Finally, we draw concluding remarks and discuss potential directions for future research in Section~\ref{sec:conc}.

\section{Control Framework for Connected and Automated Vehicles in Mixed Traffic}
\label{sec:mpc}
In this section, we first present a socially cooperative optimal control framework for a CAV while interacting with a HDV in mixed traffic, and illustrate it by a vehicle merging example.

\subsection{Problem Formulation and Control Framework} 
\label{sub:framework}

We consider an interactive driving scenario that includes a CAV and a HDV indexed by $1$ and $2$, respectively.
To facilitate connectivity, we assume that a coordinator is available to collect real-time trajectories of HDV\textendash$2$ and transmit them to CAV\textendash$1$ while also storing necessary information, \eg physical parameters of the traffic scenario. 
We also consider that there is no error or delay during the communication between the vehicles and the coordinator.

The dynamics of the vehicles are described by discrete-time models.
Let $\mathbf{x}_{i,k}$ and $\mathbf{u}_{i,k}$, $i = 1,2$, be the vectors of states and control actions, respectively, at time $k \in \NN$. Then the model of vehicle $i$ is given by
\begin{equation}
\label{eq:dynamic}
\mathbf{x}_{i,k+1} = \mathbf{f}_i (\mathbf{x}_{i,k}, \mathbf{u}_{i,k}).
\end{equation}

The goal is to develop a cooperative optimal control framework for motion planning of CAV\textendash$1$ which considers the driving intention of HDV\textendash$2$.
In previous research efforts \cite{schwarting2019social,ozkan2021socially,toghi2021cooperative}, 
CAV\textendash HDV interaction was modeled as a non-cooperative Stackelberg game, in which a leader makes a decision, then a follower makes its optimal decision with respect to the leader's decision.
Generally, computing an Stackelberg equilibrium in CAV\textendash HDV interactions can be computationally expensive for real-time optimization \cite{ozkan2021socially}.
In addition, the Stackelberg game might not ideally reflect CAV\textendash HDV interaction since determining who should be the leader and the follower is not explicit in many traffic scenarios \cite{schwarting2019social}.

To overcome these issues, we consider a non-cooperative simultaneous game 
and derive a single optimal control problem 
that yields the Nash equilibrium of the game. 
In the imposed game, CAV\textendash$1$ and HDV\textendash$2$ take control actions at the same time to minimize their objective functions which include an individual (egoistic) term and a shared (cooperative) term.
The egoistic term denotes the effort of each vehicle to achieve its own driving goal represented by functions of its states and actions, whereas the cooperative term is defined as the effort to achieve a common target involving the states and actions of both CAV\textendash$1$ and HDV\textendash$2$.
Let $l_{1} \big(\mathbf{x}_{1,k+1}, \mathbf{u}_{1,k})$ and $l_{2} \big(\mathbf{x}_{2,k+1}, \mathbf{u}_{2,k})$ be the egoistic terms of the objective functions of CAV\textendash$1$ and HDV\textendash$2$, respectively, at time $k$. 
Let $l_{12} \big(\mathbf{x}_{1,k+1}, \mathbf{u}_{1,k}, \mathbf{x}_{2,k+1}, \mathbf{u}_{2,k} \big)$ be the cooperative term of their objective function at time $k$. 
Note that $l_1$, $l_2$, and $l_{12}$ are composed appropriately through some weights to avoid one dominating the other. 


The preferences of the vehicles toward the egoistic and cooperative terms can be described by social value orientation, a commonly used concept in psychology that has been recently employed in autonomous driving research \cite{schwarting2019social,buckman2019sharing,crosato2021human,ozkan2021socially,larsson2021pro,toghi2021cooperative}.
Let $\phi_1$ and $\phi_2$ be the SVO angles for CAV\textendash$1$ and HDV\textendash$2$, respectively. 
The SVO angle is usually constrained between $0$ and $\frac{\pi}{2}$.
If the angle is equal to $0$, it means that the vehicle is purely egoistic, \ie it makes decisions that only benefit its own objective.
In contrast, if the angle is equal to $\frac{\pi}{2}$ it implies that this is a purely altruistic vehicle, \ie it optimizes the cooperative objective without concerning its own objective.
Since in reality, most drivers have at least a minimal level of egoism and altruism, we consider that $0 < \phi_1,\phi_2 < \frac{\pi}{2}$.

The objective of each vehicle in the imposed game can be formed as a weighted sum of the egoistic and cooperative terms in which the weights are determined by trigonometric functions of the SVO angles.
In particular, given the SVO angles $\phi_1$ and $\phi_2$ 
the objective function of CAV\textendash$1$ and HDV\textendash$2$ in the game is
\begin{equation}
\label{eq:uti:av}
\begin{multlined}
\bar{l}_1 (\mathbf{x}_{1,k+1}, \mathbf{u}_{1,k}, \mathbf{x}_{2,k+1}, \mathbf{u}_{2,k}) 
= l_{1} (\mathbf{x}_{1,k+1}, \mathbf{u}_{1,k}) \cos\phi_{1} \\
+ l_{12} (\mathbf{x}_{1,k+1}, \mathbf{u}_{1,k}, \mathbf{x}_{2,k+1}, \mathbf{u}_{2,k}) \sin \phi_1,
\end{multlined}
\end{equation}
and 
\begin{equation}
\label{eq:uti:hdv}
\begin{multlined}
\bar{l}_2 (\mathbf{x}_{1,k+1}, \mathbf{u}_{1,k}, \mathbf{x}_{2,k+1}, \mathbf{u}_{2,k}) 
= l_{2} (\mathbf{x}_{2,k+1}, \mathbf{u}_{2,k}) \cos\phi_2 \\
+ l_{12} (\mathbf{x}_{1,k+1}, \mathbf{u}_{1,k}, \mathbf{x}_{2,k+1}, \mathbf{u}_{2,k}) \sin \phi_2,
\end{multlined}\end{equation}
respectively.


In what follows, to simplify the notation we drop the time index $k$ whenever this does not cause any ambiguity.

To find a Nash equilibrium for the simultaneous game described above, we propose an objective function recasting the game as a potential game \cite{monderer1996potential}. 
In potential games, a Nash equilibrium at which all players minimize their own objective functions can be obtained by alternatively minimizing a single function called the potential function.
For the game between CAV\textendash$1$ and HDV\textendash$2$ where their objectives are given in \eqref{eq:uti:av} and \eqref{eq:uti:hdv}, the potential objective function can be derived as follows
\begin{equation}
\label{eq:mpc-obj}
\begin{multlined}
l (\mathbf{x}_1, \mathbf{u}_1, \mathbf{x}_2, \mathbf{u}_2) 
= l_{1} (\mathbf{x}_1, \mathbf{u}_1) \cos\phi_1 \sin \phi_2 \\
+ l_{2} (\mathbf{x}_2, \mathbf{u}_2) \sin\phi_1 \cos \phi_2 \\
+ l_{12} (\mathbf{x}_1, \mathbf{u}_1, \mathbf{x}_2, \mathbf{u}_2) \sin \phi_1 \sin \phi_2.
\end{multlined}
\end{equation}
Note that in \eqref{eq:mpc-obj} the weights for the objective functions $l_1$, $l_2$, and $l_{12}$ are determined by a weighting strategy using trigonometric functions of the SVO angles $\phi_1$ and $\phi_2$.
In the following theorem, we prove that a Nash equilibrium can be obtained by minimizing \eqref{eq:mpc-obj}.

\begin{theorem}
Consider the simultaneous game between CAV\textendash$1$ and HDV\textendash$2$ whose objective functions are given by \eqref{eq:uti:av} and \eqref{eq:uti:hdv}.
Then the minimum of the proposed objective function \eqref{eq:mpc-obj} is a Nash equilibrium of the game.
\end{theorem}



\proof
Given the vehicle dynamics in \eqref{eq:dynamic}, the functions $l$, $l_1$, $l_2$, and $l_{12}$ can be expressed as the functions of control actions $\mathbf{u}_1$ and $\mathbf{u}_2$.
Hence, we use $l ( \mathbf{u}_1, \mathbf{u}_2)$, $l_1 ( \mathbf{u}_1)$, $l_2 ( \mathbf{u}_2)$, and $l_{12} ( \mathbf{u}_1, \mathbf{u}_2)$ for brevity.
Denote $\mathbf{u}_1^*$ and $\mathbf{u}_2^*$ as the control actions that minimizing the objective function \eqref{eq:mpc-obj}, \ie
\begin{equation}
l (\mathbf{u}_1^*, \mathbf{u}_2^*) \le l (\mathbf{u}_1, \mathbf{u}_2^*), \, \forall \mathbf{u}_1,
\end{equation}
which, from \eqref{eq:mpc-obj}, is equivalent to 
\begin{equation}
\label{eq:proof-eq1}
\begin{multlined}  
l_1 (\mathbf{u}_1^*) \cos \phi_1 \sin \phi_2 + l_2 (\mathbf{u}_2^*) \sin \phi_1 \cos \phi_2 \\
+ l_{12} (\mathbf{u}_1^*, \mathbf{u}_2^*) \sin \phi_1 \sin \phi_2 \\ 
\le l_1 (\mathbf{u}_1) \cos \phi_1 \sin \phi_2 + l_2 (\mathbf{u}_2^*) \sin \phi_1 \cos \phi_2 \\
+ l_{12} (\mathbf{u}_1, \mathbf{u}_2^*) \sin \phi_1 \sin \phi_2, \, \forall \mathbf{u}_1 .
\end{multlined}
\end{equation}

Since 
$\phi_2 > 0$, from \eqref{eq:proof-eq1} we obtain
\begin{equation}
\begin{multlined}
l_1 (\mathbf{u}_1^*) \cos \phi_1 + l_{12} (\mathbf{u}_1^*, \mathbf{u}_2^*) \sin \phi_1 \\
\le l_1 (\mathbf{u}_1) \cos \phi_1 + l_{12} (\mathbf{u}_1, \mathbf{u}_2^*) \sin \phi_1, \, \forall \mathbf{u}_1,
\end{multlined}
\end{equation}
or 
\begin{equation}
\label{eq:nash-1}
\mathbf{u}_1^* = \underset{\mathbf{u}_1}{\argmin} \; l_1 (\mathbf{u}_1) \cos \phi_1 + l_{12} (\mathbf{u}_1, \mathbf{u}_2^*) \sin \phi_1.
\end{equation}

Similarly, we have
\begin{equation}
\label{eq:nash-2}
\mathbf{u}_2^* = \underset{\mathbf{u}_2}{\argmin} \; l_2 (\mathbf{u}_2) \cos \phi_2 + l_{12} (\mathbf{u}_1^*, \mathbf{u}_2) \sin \phi_2.
\end{equation}

From \eqref{eq:nash-1} and \eqref{eq:nash-2}, the result follows. \endproof

In our socially cooperative optimal control framework for motion planning of CAV\textendash$1$, we formulate a finite-time optimal control problem over a control horizon of length $H \in \NN \setminus \{0\}$. 
For simplicity, we consider constant SVO angles over the current control horizon at each time step.
Let $t$ be the current time step and $\III_t = \{ t, \dots, t+H-1 \}$ be the set of all time steps in the control horizon at time step $t$.
The finite-horizon optimal control problem for CAV\textendash$1$ is given by
\begin{subequations}
  \label{eq:mpc}
  \begin{align}
    &   
    \begin{multlined}
    \underset{ \{\mathbf{u}_{1,k}, \mathbf{u}_{2,k}\}_{k \in \III_t} }{\minimize} \quad \sum_{k \in \III_t} l (\mathbf{x}_{1,k+1}, \mathbf{u}_{1,k}, \mathbf{x}_{2,k+1}, \mathbf{u}_{2,k}) 
    \end{multlined}
    \label{eq:mpc:obj}\\
    & \text{subject to:} \nonumber  \\
    & \quad \text{\eqref{eq:dynamic}},\; \, i = 1,2, \label{eq:mpc-dyn} \\
    & \quad g_j (\mathbf{x}_{1,k+1}, \mathbf{u}_{1,k}, \mathbf{x}_{2,k+1}, \mathbf{u}_{2,k}) \le 0,\, \forall j \in \JJJ_{\text{ieq}}, \label{eq:mpc-ineq} \\ 
    & \quad h_j (\mathbf{x}_{1,k+1}, \mathbf{u}_{1,k}, \mathbf{x}_{2,k+1}, \mathbf{u}_{2,k}) = 0, \, \forall j \in \JJJ_{\text{eq}}, \label{eq:mpc-eq} 
  \end{align}
\end{subequations} 
where \eqref{eq:mpc-dyn}\textendash\eqref{eq:mpc-eq} hold for all $k \in \III_t$.
The constraints \eqref{eq:mpc-ineq} and \eqref{eq:mpc-eq} are inequality and equality constraints for system states and inputs with $\JJJ_{\text{ieq}}$ and $\JJJ_{\text{eq}}$ are sets of inequality and equality constraint indices, respectively.

\begin{remark}
The optimization variables of \eqref{eq:mpc} consist of not only variables of CAV\textendash$1$ but also of HDV\textendash$2$ so that the proposed control framework can predict HDV\textendash$2$'s states over the next control horizon given the control objective that best describe driving behavior of the human driver.
The individual objective function of HDV\textendash$2$ can be recovered from historical human driving data through offline inverse reinforcement learning \cite{kuderer2015learning}, \cite{wu2020efficient}.
If historical data are not available, the objective function can be predefined to represent rational human objectives in specific driving scenarios.
\end{remark}

The optimal control problem \eqref{eq:mpc} is implemented in a receding horizon control manner at every time step.
By estimating the SVO angle $\phi_2$ online, \eg using the method presented in Section~\ref{sec:rhe}, we can inspect the level of altruism of the human driver 
then adapt the SVO angle of CAV\textendash$1$ accordingly. 
For example, in this paper, the SVO angle $\phi_1$ of CAV\textendash$1$ is chosen as
\begin{equation}
\label{eq:svo:av}
\phi_1 = \frac{\pi}{2} - \phi_2.
\end{equation}

The justification for \eqref{eq:svo:av} is that if HDV\textendash$2$ is more egoistic then CAV\textendash$1$ needs to be more altruistic, and vice versa.

\subsection{Illustrative Example} 
\label{sub:example}

\begin{figure}[!t]
\centering
    \includegraphics[scale=0.3, bb = 150 60 850 480, clip=true]{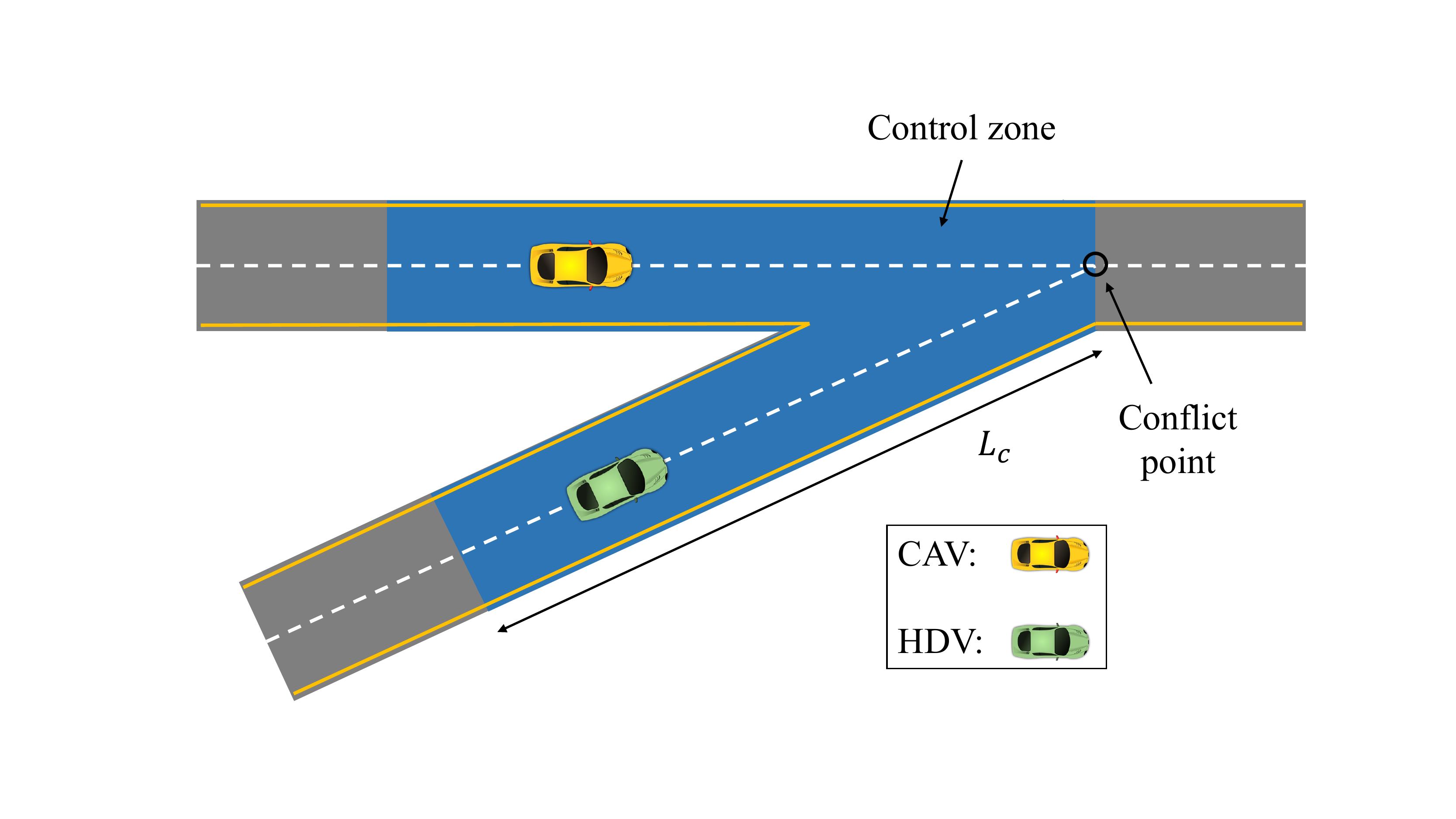}
    \caption{A merging scenario with a CAV and a HDV.}
    \label{fig:merging}
    \vspace{-20pt}
\end{figure}

To illustrate the presented control framework, we consider an example of a vehicle merging scenario illustrated in Fig.~\ref{fig:merging}. 
The area that a lateral collision can occur is called a \textit{conflict point.}
We consider a \emph{control zone} that is constituted by areas of length $L_c \in \RRplus$ upstream of the conflict point in each road (Fig.~\ref{fig:merging}).
Inside the control zone, the vehicles can communicate with a coordinator.
Note that CAV\textendash$1$ is controlled by the proposed cooperative control framework only within the control zone, whereas outside the control zone, CAV\textendash$1$ can use a speed profile given by a car-following model. 


We consider that the dynamics of the vehicles in this example can be described by the following double-integrator longitudinal dynamics
\begin{equation}
\label{eq:integrator}
\begin{split}
p_{i,k+1} &= p_{i,k} + \Delta T v_{i,k} + \frac{1}{2} \Delta T^2 a_{i,k}, \\
v_{i,k+1} &= v_{i,k} + \Delta T a_{i,k}, \\
\end{split}   
\end{equation}
where $\Delta T \in \RRplus$ is the sampling time, $p_{i,k}$ is the longitudinal position of the vehicle with respect to the conflict point at time $k$, and $v_{i,k}$ and $a_{i,k}$ are the speed and acceleration of the vehicle $i$ at time $k$, respectively.
The state and control input of vehicle $i$ are denoted by $\mathbf{x}_{i,k} = [p_{i,k}, v_{i,k}]^{\top}$ and $u_{i,k} = a_{i,k}$, respectively. 

In this example, the egoistic term for each vehicle includes minimizing both the control input 
and the deviation from the maximum allowed speed to cross the merging zone, i.e.,
\begin{equation}
\label{eq:ex-ego1}
l_{1} (\mathbf{x}_{1,k+1}, u_{1,k}) 
= w_1 a_{1,k}^2 + w_2 (v_{1,k+1} - v_{\text{max}})^2, 
\end{equation}
and 
\begin{equation}
\label{eq:ex-ego2}
l_{2} (\mathbf{x}_{2,k+1}, u_{2,k}) 
=  w_3 a_{2,k}^2 + w_4 (v_{2,k+1} - v_{\text{max}})^2,
\end{equation}
where $w_1$, $w_2$, $w_3$, and $w_4$ are positive normalized weights.
In this example, the cooperative term takes the form of a penalty function corresponding to a collision avoidance constraint \cite{cao2015cooperative} 
as follows
\begin{equation}
\label{eq:ex-cooperative}
l_{12} (\mathbf{x}_{1,k+1}, u_{1,k}, \mathbf{x}_{2,k+1}, u_{2,k})  
= \frac{w_5}{p_{1,k+1}^2 + p_{2,k+1}^2 - r^2},
\end{equation}
where $w_5$ is a positive normalized weight and $r \in \RRplus$ is a safety threshold.
Intuitively, in this merging scenario, egoistic human drivers accelerate to cross the merging zone quickly, while altruistic drivers slow down and let the other vehicle cross the merging zone prior to them. 

Next, we impose the following state and control constraints for CAV\textendash$1$, 
\begin{equation}
\label{eq:bound}
v_{\text{min}} \le v_{1,k+1} \le v_{\text{max}},\quad u_{\text{min}} \le a_{1,k} \le u_{\text{max}},
\end{equation}
for all $k \in \III_t$, where $u_{\text{min}}$, $u_{\text{max}}$ are the minimum deceleration and maximum acceleration, respectively, and $v_{\text{min}}$, $v_{\text{max}}$ are the minimum and maximum speed limits, respectively.
Note that we do not consider state and input constraints for HDV\textendash$2$ since those constraints can be violated by the behavior of human drivers. 

The receding horizon control problem for CAV\textendash$1$ in this example is thus formulated as follows
\begin{subequations}
  \label{eq:ex-mpc}
  \begin{align}
    &   
    \underset{ \{u_{1,k}, u_{2,k}\}_{k \in \III_t} }{\minimize} \quad \sum_{k \in \III_t} l (\mathbf{x}_{1,k+1}, u_{1,k}, \mathbf{x}_{2,k+1}, u_{2,k}) 
    \label{eq:ex-mpc:obj}\\
    & \text{subject to:} \nonumber  \\
    & \quad \text{\eqref{eq:integrator}},\; \forall k \in \III_t, \, i = 1,2, \label{eq:ex-mpc-dyn} \\
    & \quad \text{\eqref{eq:bound}},\; \forall k \in \III_t. \label{eq:ex-mpc-bound} 
  \end{align}
\end{subequations}

\section{Inverse Reinforcement Learning-Based Estimation for Social Value Orientation}
\label{sec:rhe}

In this section, we present a moving horizon estimation method for the SVO angle $\phi_2$ of HDV\textendash$2$ using the maximum entropy inverse reinforcement learning.

\subsection{Inverse Reinforcement Learning} 

Inverse reinforcement learning (IRL) is a machine learning technique developed to learn the underlying objective or reward of an agent by observing its behavior \cite{ziebart2008maximum}.
Using IRL, we can estimate the SVO angle of HDV\textendash$2$ that best fits predicted trajectories to actual trajectories.
Recall that in Section~\ref{sub:framework} the control input $\mathbf{u}_{2,k}^*$ of HDV\textendash$2$ at each time step $k$ is the solution of the following problem
\begin{equation}
\label{eq:mpc-hdv}
  \mathbf{u}_{2,k}^* = \;
  \underset{\mathbf{u}_{2,k}}{\argmin} \; \bar{l}_{2} = l_{2} \cos\phi_2 + l_{12} \sin\phi_2,
\end{equation}
subject to the imposed state, control, and safety constraints.

We apply the feature-based IRL approach \cite{kuderer2015learning} in \eqref{eq:mpc-hdv}, where 
we let $\mathbf{f} = [l_{2}, l_{12}]^{\top}$ and $\boldsymbol{\theta} (\phi_2) = [\cos \phi_2, \sin \phi_2]^{\top}$ be the vector of the features and the vector of weights, respectively.
Let $\RRR$ be the set of sample trajectory segments used in IRL and $\mathbf{r}_j \in \RRR$ be the $j$-th element in $\RRR$.
The goal is to find the best possible value for $\phi_2$ so that expected feature values can match observed feature values, \ie
$\EE_{p} [\mathbf{f}] = \tilde{\mathbf{f}},$
where $\tilde{\mathbf{f}}$ is the vector of average observed feature values, and $\EE_{p} [\mathbf{f}]$ denotes the expected feature values given a probability distribution over trajectories $p$. 
In general, there are many such probability distributions.
In this paper, we choose the maximum entropy IRL \cite{ziebart2008maximum} that utilizes an exponential family distribution 
and maximizes the entropy of the distribution,
yielding the following optimization problem
\begin{equation}
\label{eq:opt-phi}
\begin{split}
  & \underset{\phi_2}{\maximize} \sum_{\mathbf{r}_j \in \RRR} \log \, p \big(\mathbf{r}_j \,|\, \phi_2 \big) \\
  &  \text{subject to:} \quad 0 < \phi_2 < \frac{\pi}{2},
\end{split} 
\end{equation}
where 
\begin{equation}
\label{eq:probability}
p \big( \mathbf{r}_j \,|\, \phi_2 \big) = \frac{\exp \big(-\boldsymbol{\theta}^{\top} (\phi_2) \mathbf{f} (\mathbf{r}_j) \big)} {Z \big( \boldsymbol{\theta} (\phi_2) \big)},
\end{equation}
and $Z \big( \boldsymbol{\theta} (\phi_2) \big)$ is the partition function \cite{ziebart2008maximum}.
The probability distribution in \eqref{eq:probability} implies that the agent exponentially chooses the trajectory with a lower objective.
The constraint in \eqref{eq:opt-phi} is relaxed by parameterizing $\phi_2$ with the sigmoid function, \ie 
\begin{equation}
\label{eq:phi_to_psi}
\phi_2 = \frac{\pi}{2} \sigma (\psi)  
\end{equation}
where $\sigma (\psi) = \frac{1}{1+\exp (-\psi)}$ and $\psi \in \RR $,  leading to the following unconstrained optimization problem
\begin{equation}
\label{eq:opt-phi}
\underset{\psi}{\maximize} \quad \LLL(\psi) = \sum_{\mathbf{r}_j \in \RRR} \log \, p \big(\mathbf{r}_j \,|\, \psi \big) \\
\end{equation}

It is not possible to solve the optimization problem \eqref{eq:opt-phi} analytically, but the gradient $\nabla \LLL_{\psi}$ of the objective function in \eqref{eq:probability} with respect to $\psi$ can be computed using the chain rule as follows
\begin{equation}
\label{eq:grad-phi}
\begin{split}
\nabla \LLL_{\psi} 
&= \nabla \LLL_{\boldsymbol{\theta}}^{\top}
\begin{bmatrix}
-\sin \phi_2 \\
\cos \phi_2
\end{bmatrix}
\frac{\pi}{2} \sigma(\psi) \big( 1 - \sigma(\psi)\big),
\end{split}
\end{equation}
where $\nabla \LLL_{\boldsymbol{\theta}}$ is the gradient of the probability distribution \eqref{eq:probability} with respect to the vector of weights $\boldsymbol{\theta}$.
It can be shown that this gradient is the difference between the expected and the empirical feature values \cite{ziebart2008maximum}
\begin{equation}
\label{eq:grad-f}
\nabla \LLL_{\boldsymbol{\theta}} 
= \tilde{\mathbf{f}} - \EE_{p} [\mathbf{f}].  
\end{equation}

The average observed feature values $\tilde{\mathbf{f}}$ in \eqref{eq:grad-f} can be computed from an average of feature values for all training samples.
Meanwhile, it is generally impossible to exactly compute $\EE_{p} [\mathbf{f}]$.
In \cite{kuderer2012feature}, an approximation of the expected feature values was proposed to compute the feature values of the most likely trajectories, instead of computing expectations by sampling, as follows
\begin{equation}
\label{eq:approx-exp}
\EE_{p} [\mathbf{f}] 
\approx \mathbf{f} \big(\underset{\mathbf{r}}{\argmax}\; \log\, p (\mathbf{r} \,|\,  \psi) \big).
\end{equation}

Using \eqref{eq:approx-exp}, the gradient of the objective function in \eqref{eq:opt-phi} with respect to $\psi$ can be computed and used to update the estimation of $\phi_2$.
More details on maximum entropy IRL can be found in \cite{ziebart2008maximum}.

\subsection{Moving Horizon Estimation} 

We employ the IRL approach presented above in a moving horizon estimation manner to estimate the SVO angle $\phi_2$. 
Let $L \in \NN \setminus \{ 0 \}$ be the length of the estimation horizon and $t$ be the current time step. 
To simplify the notation, let $\mathbf{r}_j = ( \mathbf{x}_{1,t-j}, \mathbf{x}_{2,t-j}, \mathbf{x}_{1,t-j+1}, \mathbf{x}_{1,t-j+1}, \mathbf{u}_{1,t-j}, \mathbf{u}_{2,t-j} )$, for $j = 1,\dots, L$, be the tuples representing trajectory segments collected over the estimation horizon.
In other words, at each time step, we utilize the $L$ most recent sample trajectories to update $\phi_2$.
If the number of existing trajectory segments is less than the length of estimation horizon, CAV\textendash$1$ utilizes all existing trajectory segments for the estimation.

Given $L$ sample trajectories over the estimation horizon, the IRL procedure for estimating $\phi_2$ can be detailed as follows. 
At each time step, we initialize $\phi_2$ with the value computed at the previous time step, or with an arbitrary value between $0$ and $\frac{\pi}{2}$ if at the first time step.
Given an initialization of $\phi_2$, we evaluate the features for all training samples and compute the empirical feature vector averaged over all samples by
\begin{equation}
\label{eq:emp-fea}
\tilde{\mathbf{f}} = \frac{1}{L} \sum_{j=1}^{L} \mathbf{f} (\mathbf{r}_j).  
\end{equation}
Next, for $j$-th sample trajectory, we fix $\phi_2$, the trajectory $\{ \mathbf{x}_{1,k}, \mathbf{x}_{1,k+1}, \mathbf{u}_{1,k} \}$ of CAV\textendash$1$, and the initial condition $\mathbf{x}_{2,k}$, then find the optimized control actions of HDV\textendash$2$ $\mathbf{u}_{2,k}$ that minimize $\boldsymbol{\theta}^{\top} (\phi_2) \mathbf{f} (\mathbf{r}_j)$.
We denote the system trajectories resulted from the optimized HDV\textendash$2$'s actions as $\{ \mathbf{r}_1^{\phi_2}, \dots, \mathbf{r}_L^{\phi_2} \}$.
Next, we evaluate the features for all optimized trajectories and compute the approximated expected feature values $\tilde{\EE}_{p} [\mathbf{f}]$  by
\begin{equation}
\label{eq:exp-fea}
\tilde{\EE}_{p} [\mathbf{f}]
= \frac{1}{L} \sum_{j=1}^{L} \mathbf{f} (\mathbf{r}_j^{\phi_2}).
\end{equation}

Using the empirical and the expected feature values in \eqref{eq:emp-fea} and \eqref{eq:exp-fea}, we can compute the gradient of the objective function in \eqref{eq:opt-phi} with respect to $\phi_2$ by \eqref{eq:grad-phi} and \eqref{eq:grad-f}, which can be used to update $\phi_2$ by gradient ascent method as follows
\begin{equation}
\label{eq:pgd}
\psi \leftarrow \psi + \eta \nabla \LLL_{\psi},
\end{equation} 
where $\eta \in \RRplus$ is the learning rate. 

The IRL-based MHE for the SVO angle $\phi_2$ at each time step is also summarized in Algorithm~\ref{alg:IRL}.

\begin{remark}
\label{rmk:imp}
In Algorithm~\ref{alg:IRL}, at each time step, the estimation of $\phi_2$ is updated once to limit the execution time of the algorithm in real-time applications.
However, the algorithm can be repeated multiple times or until convergence.
\end{remark}



\begin{algorithm}[t]
\caption{IRL-based MHE for SVO}
\label{alg:IRL}
\begin{algorithmic}[1]
  \Require $L \in \NN \setminus \{ 0 \}$, $\eta \in \RRplus$. 
  \State Initialize $\psi$ or re-use the previous estimate.
  \State Collect sample trajectories $\mathbf{r}_j$, $j = 1,\dots,L$ over the estimation horizon.
  \For {$j = 1, \dots, L$}
  \State Compute $\mathbf{f}(\mathbf{r}_j)$ for the sample trajectory $\mathbf{r}_j$.
  \EndFor
  \State Compute the empirical feature vector $\tilde{\mathbf{f}}$ by \eqref{eq:emp-fea}.
  \For {$j = 1, \dots, L$}
  \State Find the optimized trajectory $\mathbf{r}_j^{\phi_2}$ with respect to $\phi_2$.
  \State Compute $\mathbf{f}(\mathbf{r}_j^{\phi_2})$ for the optimized trajectory $\mathbf{r}_j^{\phi_2}$.
  \EndFor
  \State Compute approximated expected feature vector $\tilde{\EE}_{p} [\mathbf{f}]$ by \eqref{eq:exp-fea}.
  \State Update $\psi$ by \eqref{eq:pgd} and $\phi_2$ by \eqref{eq:phi_to_psi}.
  \State \textbf{return $\phi_2$} 
\end{algorithmic}
\end{algorithm} \setlength{\textfloatsep}{0.2cm}

\section{Simulations}
\label{sec:sim}

In this section, we show simulation results for the vehicle merging example presented in Section~\ref{sub:example} to demonstrate the performance of the proposed framework.

\subsection{Simulation Setup} 

We simulate the behavior of HDV\textendash$2$ by imitating driving behavior of a realistic human driver from Next Generation Simulation (NGSIM) dataset \cite{ngsim}. 
Using the IRL technique, we learn an offline machine learning model to capture human driving behavior at a highway on-ramp merging.
More details on imitating human driving behavior via IRL can be found in \cite{kuderer2015learning,wu2020efficient}.
The learned human driving model is used to generate actions of HDV\textendash$2$ in the control zone of length \SI{120}{m}, while outside the control zone the speed of HDV\textendash$2$ is assumed to be constant.
By changing the weights of the obtained objective functions in the IRL human driving model, we can replicate  behavior of more egoistic or altruistic human drivers to validate the methods with different human driving styles.
The weights of the objective functions and other parameters of the RHC and MHE are given in Table~\ref{tab:sim-params}.
%
We used Python for the implementation
and CasADi \cite{andersson2019casadi} along with the built-in IPOPT solver 
for formulating and solving the optimization problems, respectively.


\begin{table}[!bt]
  \caption{Parameters of the simulations}
  \label{tab:sim-params} 
  \centering
  \begin{tabular}{p{0.085\textwidth}p{0.09\textwidth}p{0.085\textwidth}p{0.09\textwidth}}
    \toprule[1pt]
    \textbf{Parameters} & \textbf{Value} & \textbf{Parameters} & \textbf{Value} \\
    \midrule[0.5pt] 
    $w_1$ & 1 & $w_2$ & 5\\
    $w_3$ & 1 & $w_4$ & 5\\
    $w_5$ & $10^7$ & $T$ & \SI{0.1}{s}\\
    $v_{\mathrm{min}}$ & \SI{0}{m/s} & $v_{\mathrm{max}}$ & \SI{30}{m/s}\\
    $u_{\mathrm{min}}$ & \SI{-10}{m/s^2} & $u_{\mathrm{max}}$ & \SI{5}{m/s^2}\\
    $H$ & 20 & $L$ & 20\\
    $\eta$ & 1.0 & & \\
    \bottomrule[1pt] 
  \end{tabular}
\end{table}

\subsection{Results and Discussions} 

\begin{figure}[tb]
\begin{subfigure}{.48\textwidth}
\centering
\includegraphics[scale=0.44, bb = 0 0 600 340, clip=true]{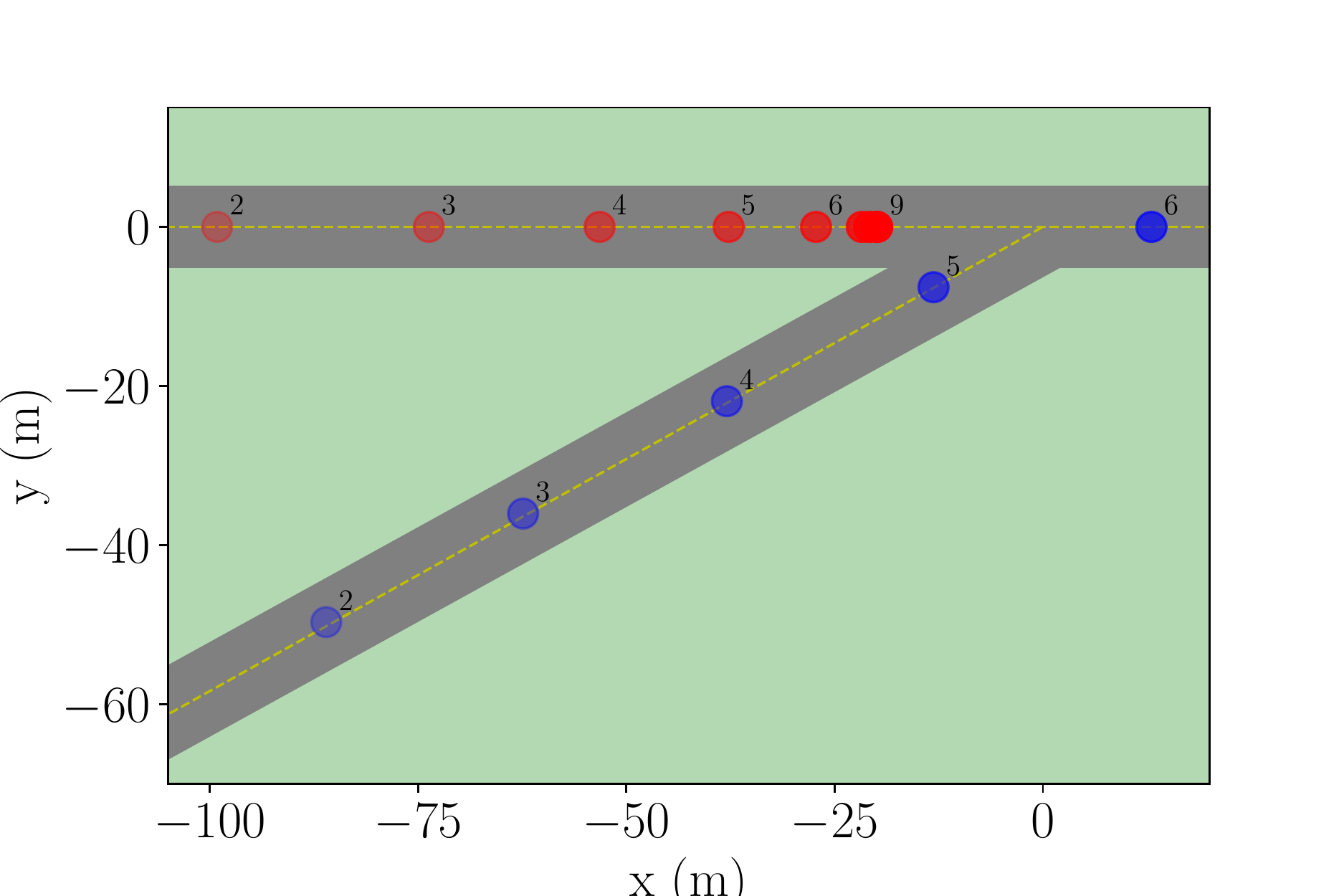}
\caption{Simulation with an egoistic human driver}
\label{fig:sim1_visual}
\end{subfigure}
\begin{subfigure}{.48\textwidth}
\centering
\includegraphics[scale=0.44, bb = 0 0 600 340, clip=true]{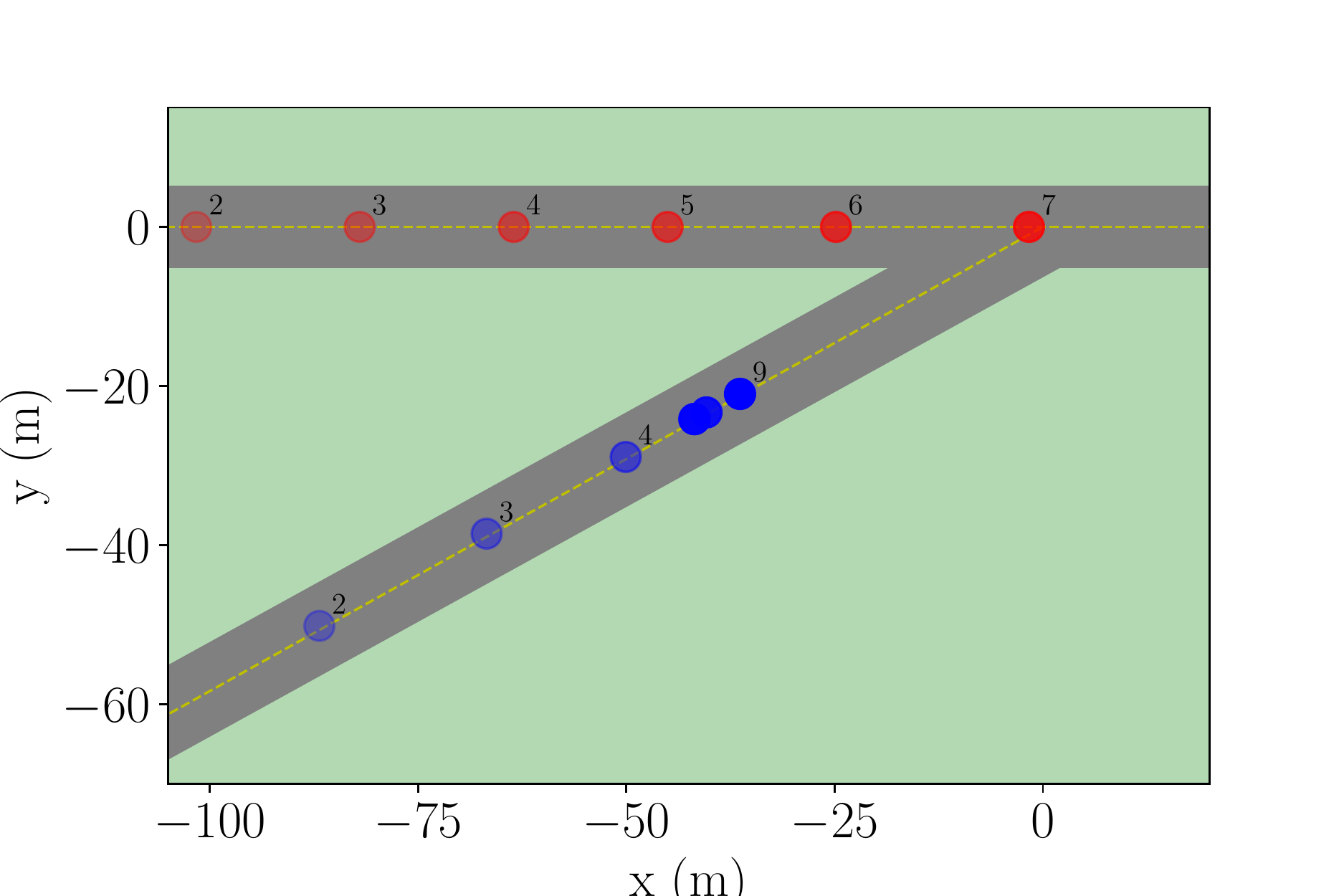}
\caption{Simulation with an altruistic human driver}
\label{fig:sim2_visual}
\end{subfigure}
\caption{Snapshots of the merging scenario with vehicle positions at every 1 second in two simulations. The CAV and the HDV are denoted by red and blue dots, respectively.}
\label{fig:sim_visual}
\end{figure}

We evaluate the performance of the proposed method in two simulation scenarios where CAV\textendash$1$ correspondingly deals with egoistic and altruistic drivers.
Figure~\ref{fig:sim1_visual} shows a snapshot of the merging scenario with positions of the vehicles at every $1$ second 
for the simulation with the egoistic driver. 
As can be seen from the figure, the egoistic HDV (blue dot) accelerates while approaching the merging zone, thus CAV\textendash$1$ (red dot) slows down to let HDV\textendash$2$ cross the merging zone first.
The results for the second simulation, in which the human driver is more altruistic, are shown in Figure~\ref{fig:sim2_visual}. 
In this case, HDV\textendash$2$ decelerates and then stops when moving close to the conflict point to give way to CAV\textendash$1$.
By noticing that the human driver is highly cooperative, CAV\textendash$1$ then speeds up to pass the merging zone prior to HDV\textendash$2$.
Using the proposed socially cooperative optimal control framework, CAV\textendash$1$ behaves differently depending on human driving preferences.
More simulation results can be found at the supplemental site, \url{https://sites.google.com/view/ud-ids-lab/socially-cooperative-control}.

\section{Conclusions}
\label{sec:conc}
In this paper, we presented a socially cooperative optimal control framework for a CAV to efficiently and safely interact with a HDV in mixed traffic.
Using the SVO angles of the vehicles, we synthesized an objective function that allows the CAV to compensate appropriately given the level of cooperation of the human driver.
The SVO of the HDV was estimated online using inverse reinforcement learning-based moving horizon estimation. 
The performance of the proposed method was validated by a vehicle merging example. 
Future work should focus on extending this framework to consider the interactions between multiple CAVs and HDVs. 
Another direction for future research should investigate the practical effectiveness of the proposed approach using experiments in a scaled robotic environment \cite{chalaki2021CSM}. 


\bibliographystyle{IEEEtran}
\bibliography{IEEEabrv,references,references_IDS}



\balance




\end{document}